\documentclass{ws-procs9x6}

\def\6#1{{\underline{#1}}}
\def\m6#1{{\underline{#1}\,}}

\newdimen\Tdim
\def\ispan{{\setbox0=\hbox{i}%
\Tdim\ht0\advance\Tdim\dp0\rule[-\dp0]{0pt}{\Tdim}}}
\def\jspan{{\setbox0=\hbox{j}%
\Tdim\ht0\advance\Tdim\dp0\rule[-\dp0]{0pt}{\Tdim}}}
\def\Tspan#1{{\setbox0=\hbox{#1}%
\Tdim\ht0\advance\Tdim\dp0\advance\Tdim.55ex\rule[-\dp0]{0pt}{\Tdim}\box0}}

\def\be{\begin{eqnarray}}
\def\ben{\begin{eqnarray*}}
\def\ee{\end{eqnarray}}
\def\een{\end{eqnarray*}}
\def\Tr{{\rm Tr}}

\def\p{\partial}

\def\D{\mathcal{D}}

\def\=:{=\hspace{-.7em}\raisebox{1.1ex}{.}\hspace{.1em}\raisebox{-0.2ex}{.} }

\newcommand {\beq}{\begin{eqnarray}}
\newcommand {\eeq}{\end{eqnarray}}

\begin{document}

\title{Static Interactions of $U(N)$ non-Abelian Vortices}

\author{Minoru~Eto}

\address{
Department of Physics, University of Pisa,\\
INFN, Sezione di Pisa\\
Largo Pontecorvo, 3,   Ed. C,  56127 Pisa, Italy\\
E-mail: minoru@df.unipi.it
}

\begin{abstract}
Interactions between non-BPS non-Abelian vortices are
studied in non-Abelian $U(1) \times SU(N)$ extensions of the Abelian-Higgs model in four dimensions.
In addition to the usual type
I/II Abelian superconductors, we find other two new regimes: type I$^*$/II$^*$.
\end{abstract}

\keywords{Soliton, vortex, superconductor}

\bodymatter

\section{Introduction}

Recently, a new type of BPS vortex was found in $U(N)$ 
gauge theories \cite{Hanany:2003hp,Auzzi:2003fs}. This is called non-Abelian vortex and
carries the non-Abelian charge
${\bf C}P^{N-1}=\frac{SU(N)_{{\rm C}+{\rm F}}}{SU(N-1)_{{\rm C}+{\rm F}}\times U(1)_{{\rm C}+{\rm F}}}$.
Readers can find 
good reviews in \cite{reviews,Eto:2006pg} and references of related works therein.
In this talk
we are interested in studying interactions between {\it non-Abelian} vortices which are non-BPS. 
The non-BPS vortices are more natural than BPS ones in a sense that the BPS always requires
a fine tuning or supersymmetry. 
It is well known that ANO vortices \cite{Abrikosov:1956sx,Nielsen}  in the
type I system feel an attractive force while those in the type II model feel a repulsive force
\cite{Gustafson:2000, Jacobs:1978ch,Bettencourt:1994kf,Speight:1996px}.
Specifically we are interested in the interactions between
vortices with different internal orientations, which is the distinct feature from the ANO case \cite{1stpaper}. 

This talk is based on \cite{Auzzi:2007wj} in collaboration with R.Auzzi and W.Vince.

\section{The model}

\subsection{A fine-tuned model}
We start with non-Abelian, $U(N)$, extension of the Abelian-Higgs model
\beq
{\cal L} =
\Tr\left[
- \frac{1}{2e^2} F_{\mu\nu}F^{\mu\nu}
+ \D_\mu H (\D^\mu H)^\dagger
- { \frac{\lambda^2 \, e^2}{4}} \left(v^2 {\bf 1}_{N} - HH^\dagger \right)^2
\right].
\label{eq:Lag_NAH}
\eeq
Here, for simplicity we take the same gauge coupling $e$ for both the $U(1)$ and $SU(N)$ groups, while $\lambda^2 \,
e^2/4$ is a scalar coupling and $v$ ($>0$) determines the Higgs VEV.
$H$ is $N$ Higgs fields in the fundamental representation of $U(N)$.
The Higgs vacuum of the model is given
by $HH^\dagger = v^2 {\bf 1}_N$.
It breaks completely the gauge symmetry, although a global color-flavor locking symmetry
$SU(N)_{\rm C+F}$ is preserved
\beq
H \to U_{\rm G} H U_{\rm F},\quad U_{\rm G} = U_{\rm F}^\dagger,\quad
U_{\rm G} \in SU(N)_{\rm G},\ U_{\rm F} \in SU(N)_{\rm F}.
\eeq
The trace part $\Tr H$ is a singlet under the color-flavor group and the traceless parts are
in the adjoint representation.
The $U(1)$ and the $SU(N)$ gauge vector bosons
have both the same mass
$M_{U(1)}=M_{SU(N)}=e \, v$.
The $N^2$ real scalar fields in
$H$ are eaten by the gauge bosons and the other $N^2$ (one singlet and
the rest adjoint) have same masses
$M_{\rm s} = M_{\rm ad} = \lambda \, e \, v$.
The critical coupling $\lambda=1$ (BPS) allows an $\mathcal{N}=2$ supersymmetric extension.

\subsection{Models with general couplings}

A generalization of (\ref{eq:Lag_NAH}) is to consider
different gauge couplings, $e$ for the $U(1)$ part and $g$ for the $SU(N)$ part,
and a general quartic scalar potential
\beq
{\cal L}
= \Tr \left[ - \frac{1}{2g^2} \hat F_{\mu\nu} \hat F^{\mu\nu} - \frac{1}{2e^2} f_{\mu\nu} f^{\mu\nu} + \D_\mu H (\D^\mu
H)^\dagger \right] - V, \label{flum}
\eeq
where we have defined $\hat F_{\mu\nu} = \sum_{A=1}^{N^2-1} F_{\mu\nu}^A T_A$ and $f_{\mu\nu} = F_{\mu\nu}^0 T^0$ with
$\Tr(T^AT^B) = \delta^{AB}/2$ and $T^0 = {\bf 1}/\sqrt{2N}$
The scalar potential is:
\beq
V 
=\frac{\lambda_g^2g^2}{4} \Tr \hat X^2 + \frac{\lambda_e^2e^2}{4}\Tr \left(X^0T^0 - v^2{\bf 1}_N\right)^2,
\label{eq:flum_pot}
\eeq
where
$HH^\dagger = X^0T^0 + \hat X$ and $\hat X = 2
\sum_{A=1}^{N^2-1} \left(H^{i\dagger}T^AH_i\right)T^A$.
The symmetries is same as the previous fine-tuned model~(\ref{eq:Lag_NAH}). 
In this model, the $U(1)$ and the $SU(N)$ vector bosons have different masses
$M_{U(1)}=e \,v, \ M_{SU(N)}=g \, v.$
Moreover, the singlet part of $H$ has a mass $M_{\rm s}$ different from that of the adjoint part
$M_{\rm ad}$ as
$M_{\rm s}=\lambda_e \, e \, v, \  M_{\rm ad}=\lambda_g \, g \, v$.
For the critical values $\lambda_e=\lambda_g=1$,
the Lagrangian again allows an $\mathcal{N}=2$ susy extension.


\subsection{Vortex equations in the fine-tuned model}

Let us make the
following rescaling of fields and coordinates:
\beq
H \rightarrow v H,\quad
W_\mu \rightarrow ev W_\mu,\quad
x_\mu \rightarrow \frac{x_\mu}{ev}.
\label{eq:rescale}
\eeq
The masses of vector and scalar bosons are
rescaled to
\beq
M_{U(1)} = M_{SU(N)} = 1,\qquad
M_{\rm s} = M_{\rm ad} = \lambda.
\eeq

In order to construct non-BPS non-Abelian vortex solutions,
we have to solve the equation of motion
derived from the Lagrangian (\ref{eq:Lag_NAH}),
\beq
\D_\mu F^{\mu\nu} - \frac{i}{2}\left[H (\D^\nu H)^\dagger - (\D^\nu H) H^\dagger\right] = 0,
\label{eq:eom1}\\
\D_\mu\D^\mu H + \frac{\lambda^2}{4}\left(1-HH^\dagger\right)H = 0.
\label{eq:eom2}
\eeq
From now on, we restrict ourselves to static configurations
depending only on the coordinates $x^1,x^2$. Here we introduce a
complex notation
$z = x^1 + i x^2,
\p = \frac{\p_1-i\p_2}{2},\ 
W = \frac{W_1 - iW_2}2,\ 
\D = \frac{\D_1 - i\D_2}2 = \p + iW$.
Instead of the equation of motions itself, it might be better to study gauge invariant
quantities. For that purpose let us define
\beq
\bar W (z,\bar z) = - i S^{-1}(z,\bar z)\bar\p S(z,\bar z),\quad
H(z,\bar z) = S^{-1}(z,\bar z) \tilde H(z,\bar z),
\label{eq:decomposition}
\eeq
where $S$ takes values in $GL(N,{\bf C})$ and it is in the fundamental representation of $U(N)$ while the gauge singlet
$\tilde H$ is an $N \times N$ complex matrix. There is an equivalence relation $(S,\tilde H) \sim (V(z)S,V(z)\tilde
H)$, where $V(z)$ is a holomorphic $GL(N,{\bf C})$ matrix with respect to $z$.
The gauge group $U(N)$ and the flavor symmetry act as follows
\beq
S(z,\bar z) \to U_{\rm G} S(z,\bar z),\quad
H_0(z) \to H_0(z) U_{\rm F}.
\eeq
An important gauge invariant quantity is now constructed as
\beq
\Omega (z,\bar z) \equiv S(z,\bar z) S(z,\bar z)^\dagger.
\eeq
With respect to the gauge invariant objects, the equations of motion are
\beq
4 \bar \p^2 \left( \Omega \p \Omega^{-1} \right) - \tilde H \bar\p \left( \tilde H^\dagger \Omega^{-1} \right)
+ \bar\p \tilde H \tilde H^\dagger \Omega^{-1} = 0 ,\qquad\quad
\label{eq:eom_omega1}\\
\Omega \p \left( \Omega^{-1} \bar \p \tilde H \right)
+ \bar \p \left( \Omega \p \left( \Omega^{-1} \tilde H \right)\right)
+ \frac{\lambda^2}{4} \left( \Omega - \tilde H \tilde H^\dagger \right) \Omega^{-1} \tilde H = 0.
\label{eq:eom_omega2}
\eeq
These equations must be solved with the
boundary conditions for $k$ vortices
$\det \tilde H \rightarrow z^k,\ \Omega \rightarrow \tilde H \tilde H^\dagger$ as $z \rightarrow \infty$.

\subsection{BPS Limit}

For the later convenience, let us 
see the BPS limit $\lambda \to 1$. 
It can be done by just taking a holomorphic function $\tilde H$ with respect to $z$ as
\beq
\tilde H = H_0(z).
\label{eq:aho}
\eeq
Then the equations (\ref{eq:eom_omega1}) and (\ref{eq:eom_omega2}) reduce to the 
single matrix equation
\beq
\bar \p \left( \Omega \p \Omega^{-1} \right)
+ \frac{1}{4} \left( {\bf 1} - H_0 H_0^\dagger \Omega^{-1} \right) = 0.
\label{eq:master}
\eeq
This is the master
equation for the BPS non-Abelian vortex and the holomorphic matrix $H_0(z)$ is called the moduli matrix
\cite{Eto:2005yh,Eto:2006pg}. All the complex
parameters contained in the moduli matrix are moduli of the BPS vortices. For example, the position of the vortices can
be read from the moduli matrix as zeros of its determinant $\det H_0(z_i) = 0$. Furthermore, the number of vortices
(the units of magnetic flux of the configuration) corresponds to the degree of $\det H_0(z)$ as a polynomial with
respect to $z$. The classification of the moduli matrix for the BPS vortices is given in
Ref.~\cite{Eto:2005yh,Eto:2006pg}.

Consider $U(2)$ gauge theory. The
minimal vortex is generated by
\beq
H_0^{(1,0)} =
\left(
\begin{array}{cc}
z-z_0 & 0 \\
-b' & 1
\end{array}
\right),\qquad
H_0^{(0,1)} =
\left(
\begin{array}{cc}
1 & -b \\
0 & z-z_0
\end{array}
\right).
\label{eq:mm_single}
\eeq
$z_0$ corresponds to the position of the vortex and
$b$ and $b'$
are the internal orientation. 
One can extract the
orientation as the  null eigenvector of $H_0(z)$ at the vortex position
$z=z_0$ as
\beq
\vec\phi^{~(1,0)} =
\left(
\begin{array}{c}
1\\
b'
\end{array}
\right)
\quad\sim\quad
\vec \phi^{~(0,1)}
=
\left(
\begin{array}{c}
b\\
1
\end{array}
\right).\label{fundorient}
\eeq
Here ``$\sim$" stands for an identification up to complex non zero factors: $\vec \phi \sim \lambda \vec\phi$,
$\lambda \in {\bf C}^*$, so that we have found
${\bf C}P^1$~\cite{Eto:2005yh,Eto:2006pg}. We call
two non-Abelian vortices with equal orientational vectors {\it parallel}, while orthogonal orientational
vectors {\it anti-parallel}.

Arbitrary two vortices (the center of mass is fixed to be zero and the overall orientaion is fixed) 
is given by
\beq
H_{0 \ {\rm red}}^{(1,1)} \equiv \left(
\begin{array}{cc}
z-z_0 & -\eta\\
0 & z + z_0
\end{array}
\right).\label{reducedpatch}
\eeq
The orientational vectors are then of the form
\beq
\vec\phi^{~(1,1)}_1\big|_{z=z_0} =
\left(
\begin{array}{c}
1\\
0
\end{array}
\right),\qquad
\vec\phi^{~(1,1)}_2\big|_{z=-z_0} =
\left(
\begin{array}{c}
\eta\\
-2 z_0
\end{array}
\right).
\label{eq:oris}
\eeq

\section{Vortex interaction in the fine-tuned model}

\subsection{$(k_1,k_2)$ coincident vortices}

\begin{figure}[ht]
\begin{center}
\begin{tabular}{cc}
\begin{tabular}{c|cc}
$\lambda$ & $k=1$ & $k=2$\\
\hline
0.8 & 0.91231 & 1.77407 \\
0.9 & 0.95737 & 1.88936 \\
1   & 1.00000 & 2.00000 \\
1.1 & 1.04053 & 2.10655 \\
1.2 & 1.07922 & 2.20944 \\
\end{tabular}
&\includegraphics[width=7cm]{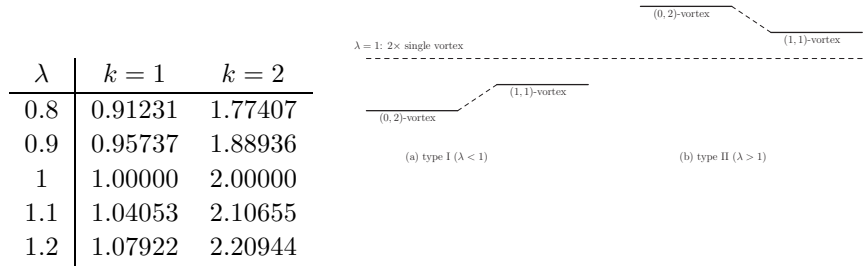}
\end{tabular}
\caption{{\small Spectrum of the $(0,2)$ and $(1,1)$ coincident vortices.}}
\label{fig:spectrum}
\end{center}
\end{figure}

The minimal winding solution in the non-Abelian gauge theory is a mere embedding of the ANO solution into the
non-Abelian theory. Embedding is also useful for another simple non-BPS configurations.
Let us start with the moduli
matrix for a configuration of  $k$ coincident vortices. The axial symmetry
allows a reasonable ansatz for $\Omega$ and $\tilde H$
\beq
\Omega^{(0,1)} =
\left(
\begin{array}{cc}
1 & 0\\
0 & w(r)
\end{array}
\right),\qquad
\tilde H^{(0,1)} =
\left(
\begin{array}{cc}
1 & 0\\
0 & f(r) z^k
\end{array}
\right).
\label{eq:mm_para}
\eeq
We call this ``$(0,k)$-vortex''.
When $k \ge 2$, it is possible that the ansatz (\ref{eq:mm_para}) does not give the true solution (minimum of the energy) of
the equations of motion (\ref{eq:eom_omega1}) and (\ref{eq:eom_omega2}). This is because there could be repulsive
forces between the vortices. With ansatz (\ref{eq:mm_para}) we fix the positions of all the vortices at the origin by
hand. The master equation (\ref{eq:master}) is nevertheless still useful to investigate the
interactions between two vortices. The results are listed in Table \ref{fig:spectrum}.
For $\lambda = 1$, the masses are identical to integer values, up to $10^{-5}$ order, which are nothing but the winding
number of the vortices.

There is another type of composite configuration which can easily be analyzed numerically
\beq
\Omega^{(1,1)} =
\left(
\begin{array}{cc}
w_1(r) & 0\\
0 & w_2(r)
\end{array}
\right),\qquad
\tilde H^{(1,1)} =
\left(
\begin{array}{cc}
f_1(r) z^{k_1} & 0\\
0 & f_2(r) z^{k_2}
\end{array}
\right).
\label{eq:mm_anti_para}
\eeq
This ansatz corresponds to a configuration with $k_1$ composite vortices which wind in the first diagonal $U(1)$ subgroup
of $U(2)$ and with $k_2$ coincident vortices that wind the second diagonal $U(1)$ subgroup.
We refer to these as a ``$(k_1,k_2)$-vortex''. The mass of a $(k_1,k_2)$-vortex is thus the
sum of the mass of the $(k_1,0)$-vortex and that of the $(0,k_2)$-vortex.

We call the non-Abelian vortices in the fine-tuned model for
$\lambda < 1$ type I, while they will be called type II for $\lambda>1$. From Fig.~\ref{fig:spectrum}, we can see that 
in the type I case, the $(0,2)$-vortex is energetically preferred to the $(1,1)$-vortex, while in type II case the
$(1,1)$-vortex is preferred. If the two vortices are separated sufficiently,
regardless of their orientations, the mass of two well separated vortices is twice that of the single vortex. This mass
is equal to the mass of the $(1,1)$-vortex.

\subsection{Effective potential for coincident vortices \label{sect:eff_coinc}}

The dynamics of BPS solitons can be investigated by the so-called moduli approximation~\cite{Manton:1981mp}. The
effective action is a massless non-linear sigma model whose target space is  the moduli space. 
If the coupling constant $\lambda$ is close to the BPS limit $\lambda = 1$, we can still use the moduli
approximation, to investigate dynamics of the non-BPS non-Abelian vortices 
by adding a potential of order $|1-\lambda^2|
\ll 1$.
To this end, we write the Lagrangian
\beq
\tilde {\cal L} = \tilde {\cal L}_{\rm BPS}+\frac{(\lambda^2 - 1)}{4} \left( {\bf 1}_{N} - HH^\dagger \right)^2.
\label{eq:Lag_Hind}
\eeq
We get non-BPS corrections of order $O(\lambda^2 - 1)$ by putting BPS solutions
into Eq.~(\ref{eq:Lag_Hind}). The energy functional thus takes the following form
\beq
{\cal E}= 2 + (\lambda^2 - 1) {\cal V},\quad
 {\cal V} = \frac{1}{8 \pi}  \int dx^1dx^2 \ \Tr \left({\bf 1} - \left|H_{\rm
BPS}(\varphi_i)\right|^2 \right)^2\label{eff gen}
\eeq
where $H_{\rm BPS}(\varphi_i)$ stands for the BPS solution. 
We have defined a reduced effective potential ${\cal V}$ which is independent of $\lambda$.
The first term corresponds to the mass of two BPS vortices and the second term is the deviation
from the BPS solutions which is nothing but the effective potential we want.

\begin{figure}[ht]
\begin{center}
\includegraphics[width=4cm]{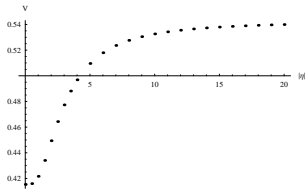}
\caption{{\small Numerical plots of the effective reduced potential ${\cal V}(|\eta|)$.}} \label{fig:num_V}
\end{center}
\end{figure}

To have the effective potential on the moduli space of coincident vortices, 
it suffices
to consider only the matrix (\ref{reducedpatch}) with turning off the relative distance $z_0$.
In order to evaluate it, we need to solve the BPS equations with
an intermediate value of $\eta$.
Because of the  axial symmetry and the
boundary condition at infinity
$\Omega  \rightarrow H_0(z)H_0^\dagger(\bar z)$,
 we can make an ansatz
\beq
\Omega^{(1,1)} = \left(
\begin{array}{cc}
w_1(r) & -\eta e^{-i\theta}w_2(r) \\
-\eta e^{i\theta}w_2(r) & w_3(r)
\end{array}
\right).
\label{eq:omega_11}
\eeq
The advantage of the moduli matrix formalism is that only three functions $w_i(r)$ 
are needed and the formalism itself is gauge
invariant. 
The effective potential can be obtained by plugging numerical solutions into Eq.~(\ref{eff gen}). 
The result is shown
in Fig.~\ref{fig:num_V}.

The type II  effective
potential has the same qualitative behavior as showed in the figure. It has a minimum at $|\eta|=0$. This
matches the previous result that the $(1,1)$-vortex is energetically preferred to the $(2,0)$-vortex. The
type I effective potential can be obtained just by flipping the overall sign of
that of the type II case. Then the effective potential always takes a negative value, which is consistent
with the fact that the masses of the type I vortices are less than that of the BPS vortices. 
Contrary to the type II case, the type I potential has a minimum at $|\eta| = \infty$, 
so that the $(2,0)$-vortex is preferred to the $(1,1)$ vortex.

\subsection{Interaction  at generic vortex separation}

\begin{figure}[ht]
\begin{center}
\begin{tabular}{ccc}
\includegraphics[width=3.5cm]{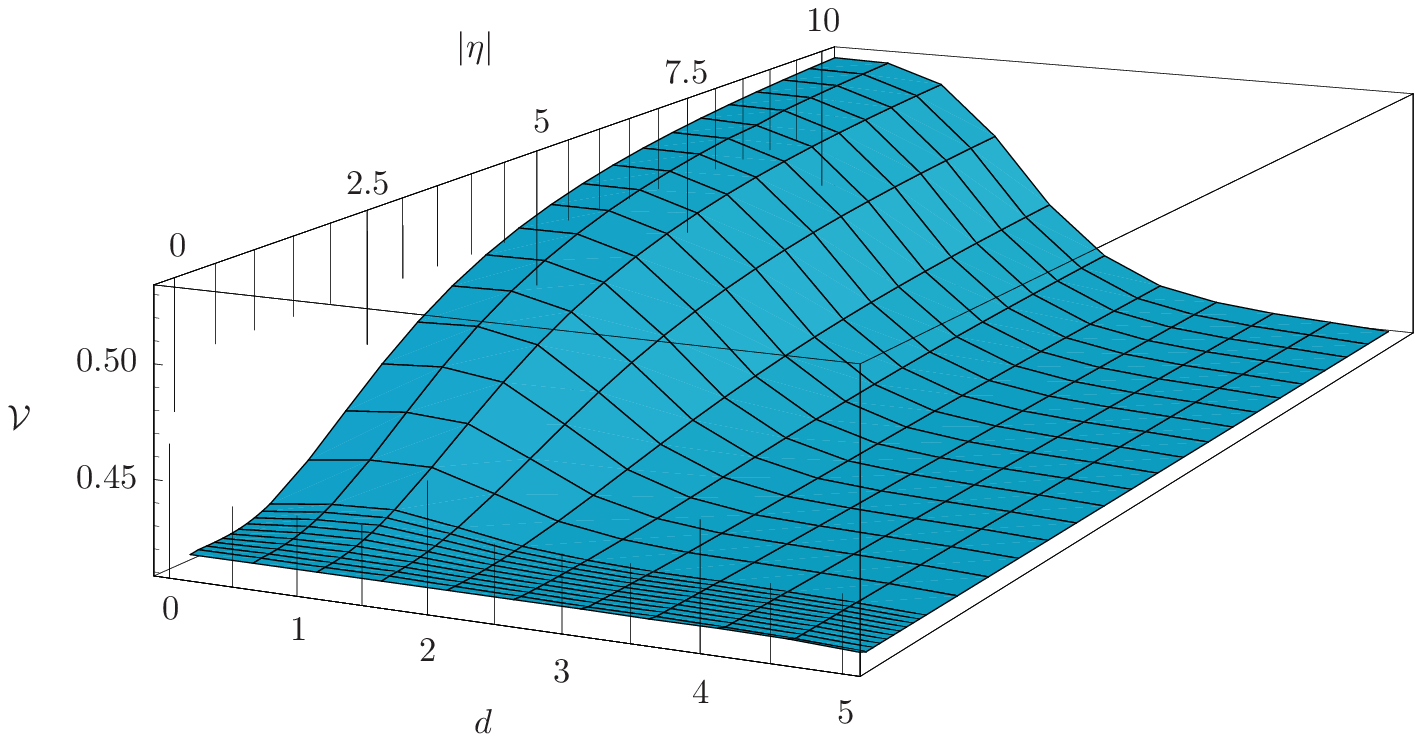}&
\includegraphics[width=3.5cm]{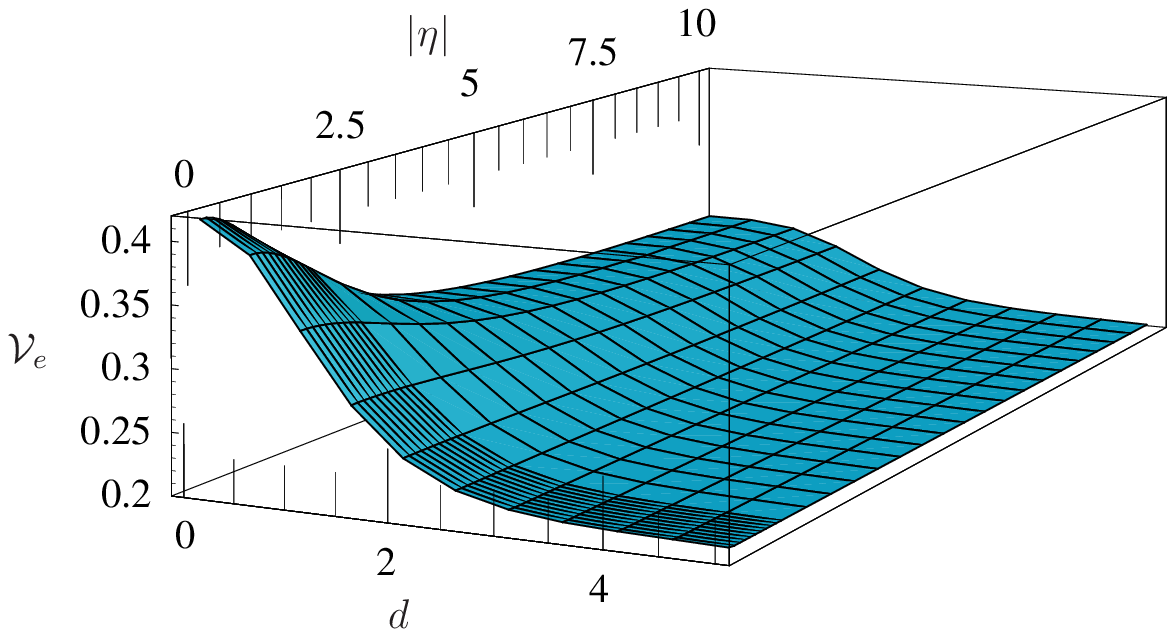} &
\includegraphics[width=3.5cm]{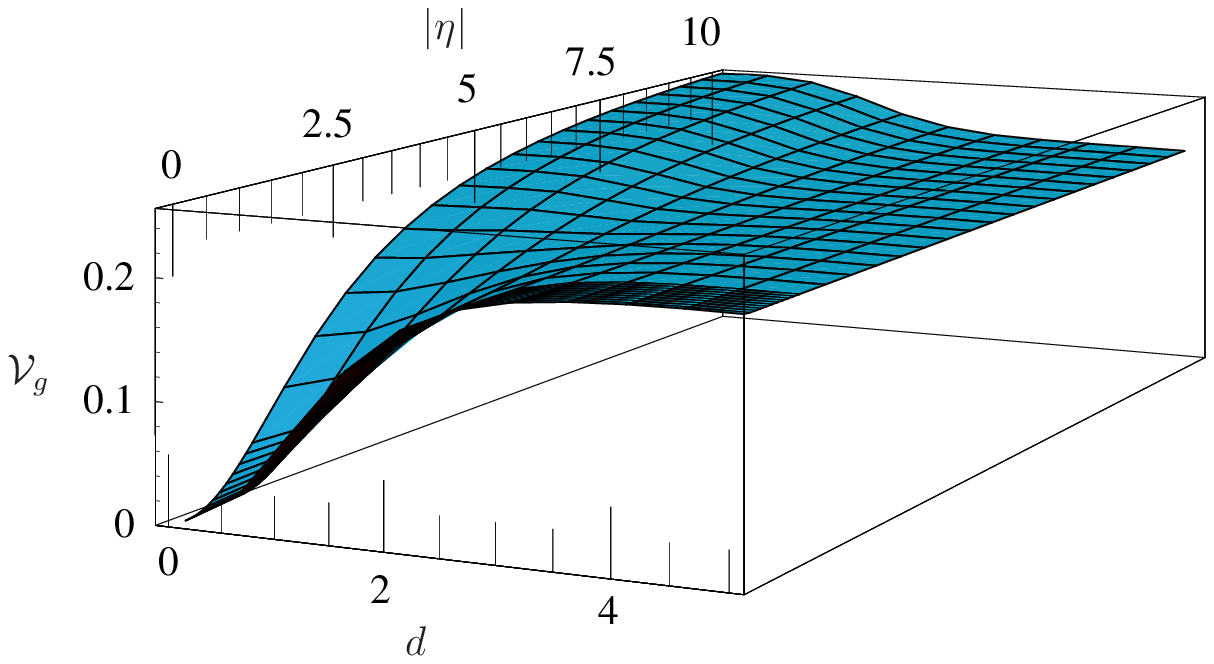}
\end{tabular}
\caption{{\small 
Left panel is the effective potential ${\cal V}(\eta,d)$.
The Abelian potential ${\cal V}_e$ (middle) and the non-Abelian potential ${\cal V}_g$ (right) for
$\gamma=1$.}}\label{fig:effv_sep}
\end{center}
\end{figure}

Next we go on investigating the interactions of non-Abelian vortices in the $U(2)$ gauge group at generic
distances. We will again use the moduli space approximation.
The generic configurations are described  by the moduli matrices in Eq.~(\ref{reducedpatch}). 
By putting the two vortices on the real axis, we can
reduce $z_0$ to a real parameter $d$.
So $2d$ is the relative distance and $\eta$ the relative orientation.
Now let us study the effective potential as function of $\eta$ and $d$.
As before, we first need the numerical solution to the BPS master
equation. 
Despite the great complexity by broken axial symmetry, the moduli matrix formalism is a powerful tool
and the relaxation method is very effective
to solve the problem. 
Once we get the numerical solution, the effective potential is
obtained by plugging them into Eq.~(\ref{eff gen}), see Fig.~\ref{fig:effv_sep}. It for
the type II has the same shape, up to a small positive factor ($\lambda^2-1$).
The potential forms a hill whose top is at $(d,|\eta|)= (0,\infty)$. It clearly shows that two vortices feel
repulsive forces, in both the real and internal space, for every distance and relative orientation. The minima of the
potential has a flat direction along the $d$-axis where the orientations are anti-parallel $(\eta=0)$ and along the $\eta$
axis at infinite distance $(d=\infty)$.
Therefore the anti-parallel vortices
do not interact.
In the type I case ($\lambda < 1$) the effective potential is upside-down of that of the type II case. There is unique
minimum of the potential at $(d,|\eta|)= (0,\infty)$. This means that attractive force works not only for the distance
in real space but also among the internal orientations.

\section{Vortices with generic couplings}

In this section we sutudy the general model
defined in Eqs.~(\ref{flum}) and (\ref{eq:flum_pot}). 
We have three effective couplings $\gamma = g/e,\lambda_e,\lambda_g$ 
after the rescaling (\ref{eq:rescale}). The masses of particles are rescaled as
\beq
M_{U(1)} = 1,\quad M_{SU(N)} = \gamma,\quad M_{\rm s} = \lambda_e,\quad M_{\rm ad} = \gamma \lambda_g.\label{masses}
\eeq
In order to find the effective potential on the moduli space as before, we
need to clarify BPS configurations.
The moduli matrix in (\ref{eq:aho}) is still valid, while
the master equation (\ref{eq:master}) get a modification
\beq
4\bar\p \left(\Omega \p \Omega^{-1} \right) = \Omega_0 \Omega^{-1} - {\bf 1}_N +
(\gamma^2 - 1) \left( \Omega_0 \Omega^{-1} - \frac{\Tr \left(\Omega_0 \Omega^{-1}\right)}{N} {\bf 1}_N \right)
\label{eq:gene_master}
\eeq
where $\Omega = SS^\dagger$ is same as before and $\Omega_0 \equiv H_0H_0^\dagger$.
It turns out that the effective potential consists of the Abelian and the non-Abelian potentials
\beq
{\cal V}_{e}(\eta,d;\gamma) = \int d\tilde x^2\ \Tr (F_{12}^0T^0)^2,\quad
{\cal V}_{g}(\eta,d;\gamma) =  \int d\tilde
x^2\ \Tr (\hat F_{12})^2. \label{potentials}
\eeq
The true potential is a linear combination of them
\beq
V(\eta,d;\gamma,\lambda_e,\lambda_g) = (\lambda_e^2 - 1) {\cal V}_e(\eta,d;\gamma) + \frac{\lambda_g^2
-1}{\gamma^2} {\cal V}_g(\eta,d;\gamma). \label{eq:gene_effpot}
\eeq

\subsection{Equal gauge coupling $\gamma=1$ revisited}

The effective potential with $\gamma=1$ and $\lambda = \lambda_g=\lambda_e$ in the left panel of
Fig.~\ref{fig:effv_sep} should be now decomposed in the two potentials, see the middle and the right panels
in Fig.~\ref{fig:effv_sep}.
In the case with $\lambda_e^2-1 > 0$ and $\lambda_g^2-1>0$, the effective potential will have
the same qualitative behaviors like the reduced potentials in the Figs.~\ref{fig:effv_sep}. 
The figures shows how ${\cal V}_e$ and ${\cal V}_g$ behaves very differently. In particular, the
Abelian potential is always repulsive, both in the real and internal space.
The non-Abelian potential is on the contrary sensitive on the
orientations.
Fig.~\ref{fig:effv_sep} shows that it is repulsive for parallel vortices while it is
attractive for anti-parallel ones. When the two scalar
couplings are equal, $\lambda_e^2=\lambda_g^2$, the two
potentials exactly cancel for anti-parallel vortices.

Of course, the true effective potential depends on $\lambda_e$ and $\lambda_g$ through the combination in
Eq.~(\ref{eq:gene_effpot}). This indicates the interaction between non-Abelian vortices is quite rich in comparison
with that of the ANO vortices.

\subsection{Different gauge coupling  $\gamma \neq 1$}

\begin{figure}[ht]
\begin{center}
\begin{tabular}{ccccc}
\includegraphics[width=3cm]{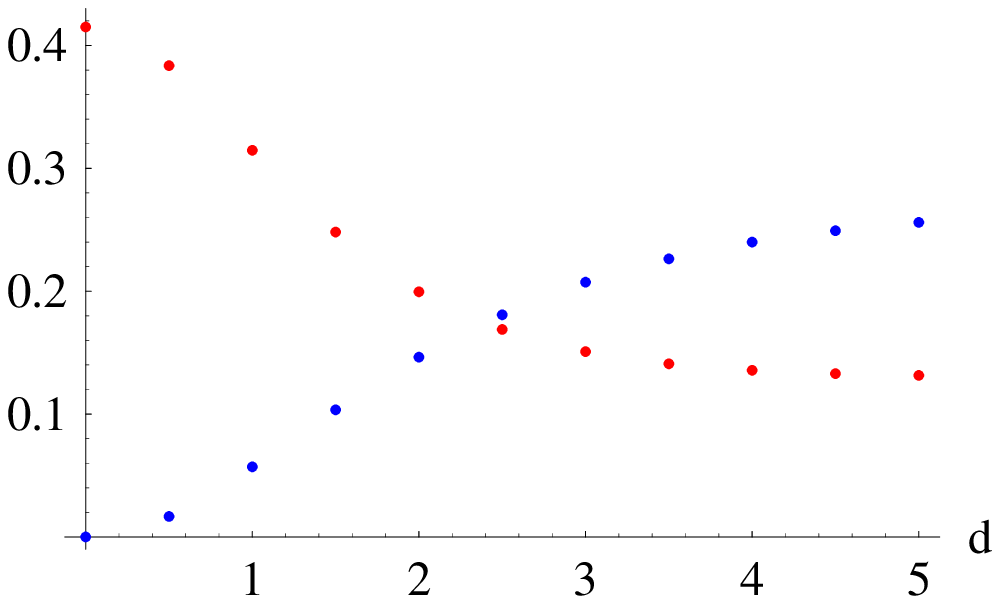}
&&
\includegraphics[width=3cm]{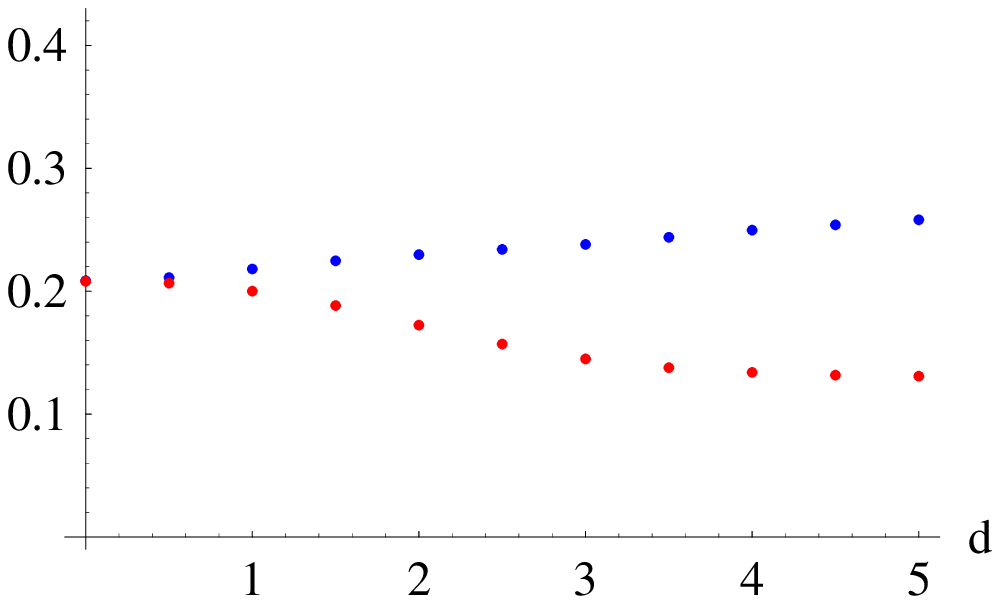}
&&
\includegraphics[width=3cm]{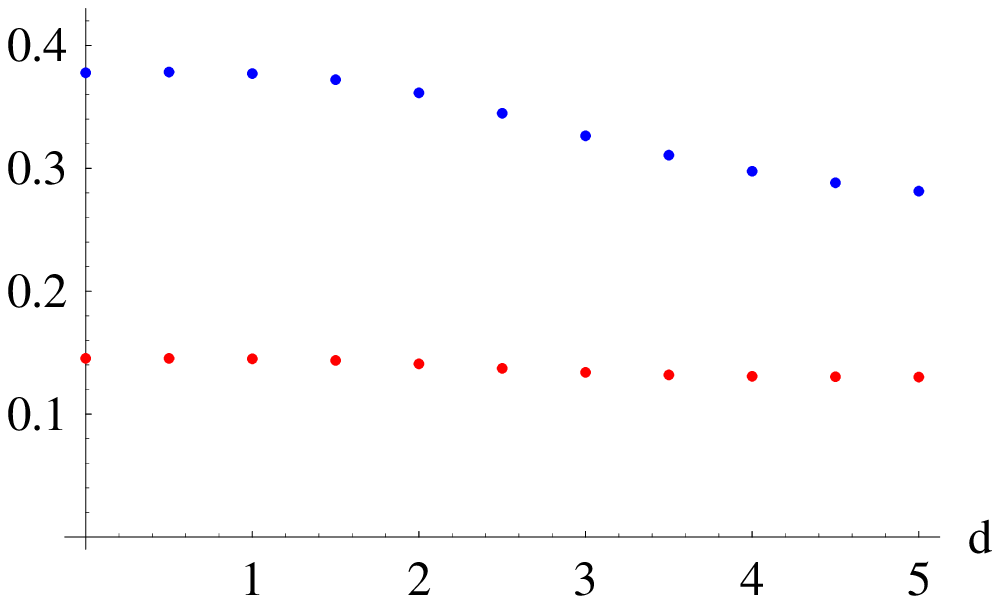}\\
anti-parallel ($\eta=0$) && intermediate ($\eta=4$) && parallel ($\eta=\infty$)
\end{tabular}
\caption{{\small Effective potential with $\gamma=1/2$ vs. separation. (red, blue) = (${\cal V}_e$, ${\cal V}_g$). }}
\label{fig:effv_e=2g}
\ \\
\begin{tabular}{ccccc}
\includegraphics[width=3cm]{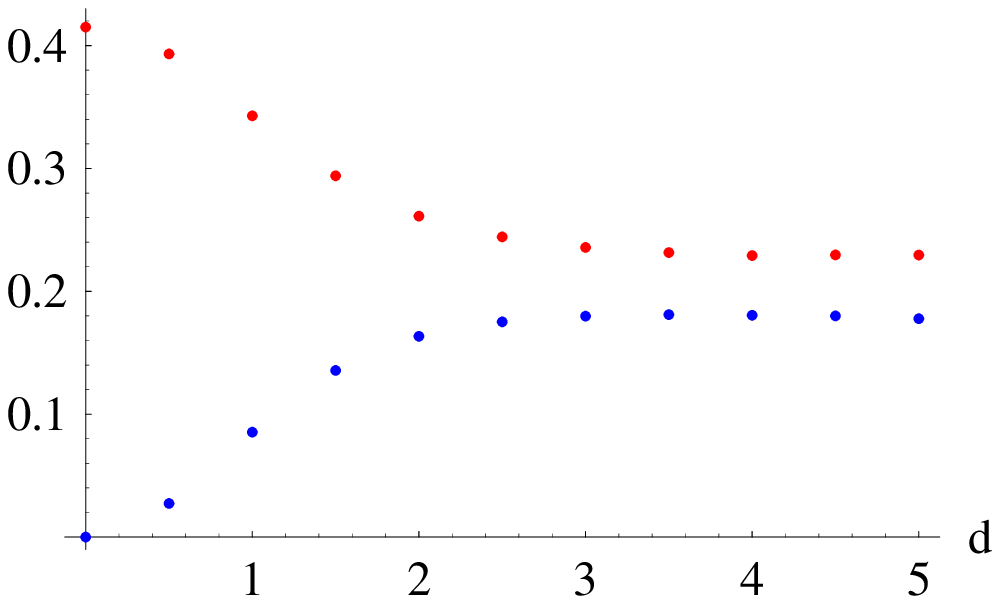}
&&
\includegraphics[width=3cm]{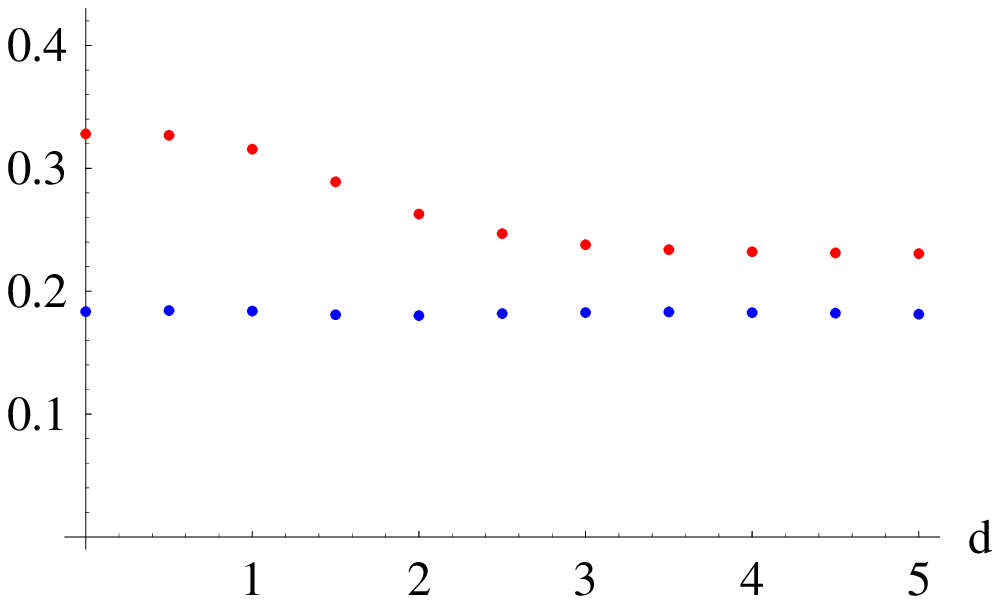}
&&
\includegraphics[width=3cm]{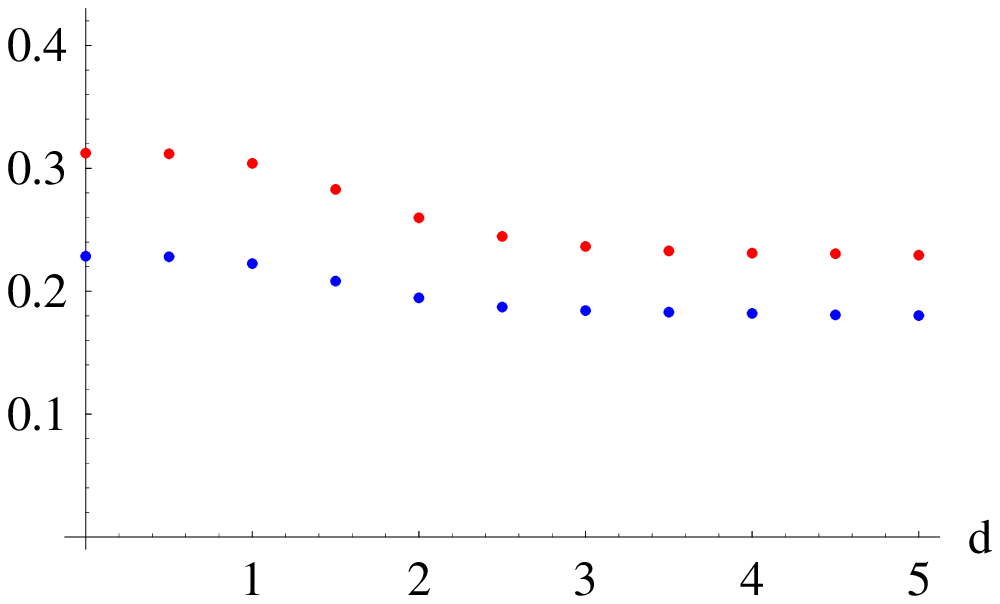}\\
anti-parallel ($\eta=0$) && intermediate ($\eta=4$) && parallel ($\eta=\infty$)
\end{tabular}
\caption{{\small Effective potential with $\gamma=1.3$ vs. separation. (red, blue) = (${\cal V}_e$, ${\cal V}_g$). }}
\label{fig:effv_g13}
\end{center}
\end{figure}

\begin{figure}[ht]
\begin{center}
\includegraphics[width=4.5cm]{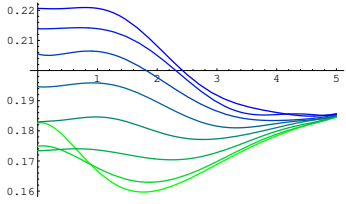}
\caption{{\small $\gamma=1/2$, $\lambda_e=1.2$, $\lambda_g=1.06$: From $\eta=0$ (green) to $\eta=7$ (blue) with
$d=0\sim5$ for each $\eta$.}}\label{bound}
\end{center}
\end{figure}

We now consider interactions between non-Abelian vortices with different gauge coupling $e\neq g$ ($\gamma \neq 1$).
In Figs.~\ref{fig:effv_e=2g} and \ref{fig:effv_g13} we show two numerical examples for the reduced effective potentials
${\cal V}_e$, ${\cal V}_g$
given in Eq.~(\ref{potentials}).
These show that the qualitative features of ${\cal V}_e$ and ${\cal V}_g$ are basically the same as
what is discussed
in the equal gauge coupling case $(\gamma=1)$.
The true effective potential in  Eq.~(\ref{eq:gene_effpot}) 
depends on the three parameters $\gamma$, $\lambda_e$ and $\lambda_g$. 
We can have potentials which develop a global minimum at some finite non zero
distance, see
Fig.~\ref{bound}
The figure shows the presence of a minimum around $d\sim
2.$ This kind of behavior
have not been found for the ANO type I/II vortices and the possibility of bounded vortices really results from the
non-Abelian symmetry.

\section{Interaction at large vortex separation}

\subsection{Vortices in fine-tuned models $e=g$ and $\lambda_e=\lambda_g$ \label{sec:ana_int_fine}}

We study an asymptotic forces between vortices
at large separation, following Refs. \cite{Speight:1996px}.
We need to find asymptotic behaviors around $(1,0)$-vortex
\beq
H_0(z)^{(1,0)} =
\left(
\begin{array}{cc}
z & 0\\
0 & 1
\end{array}
\right),\qquad
\vec\phi_1^{(1,0)} =
\left(
\begin{array}{cc}
1\\
0
\end{array}
\right).
\eeq
We are
lead to the well known asymptotic behavior of the
ANO vortex
\beq
H_{[1,1]} = \left( 1 + \frac{q}{2\pi}K_0(\lambda r) \right) e^{i\theta},\quad
\bar W_{[1,1]} = - \frac{i}{2} \left(\frac{1}{r} - \frac{m}{2\pi}K_1(r) \right) e^{i\theta}, \label{eq:asymW}
\eeq
where $K_1 \equiv - K'_0$ and
we have defined $H_{[1,1]}$ and $\bar W_{[1,1]}$ as $[1,1]$ elements of
$H$ and $\bar W$ in Eq.~(\ref{eq:decomposition}) with the $k=1$ ansatz (\ref{eq:mm_para}).

Next we treat the vortices as point particles in a linear field theory
coupled with a scalar source $\rho$ and a vector current $j_\mu$.
To linearize the Yang-Mills-Higgs Lagrangian, we choose a gauge such that
the Higgs fields is given by hermitian matrix
$H =
{\bf 1}_2
+ \frac{1}{ 2}h^i\sigma_i,\ W_\mu = \frac{1}{ 2}w_\mu^i\sigma_i
$ with $\sigma=({\bf 1}_2,\vec\sigma)$.
with all $h^a,w_\mu^a$ are real.
Then the quadratic part of the Lagrangian is
\beq
{\cal L}^{(2)}_{\rm free} = \sum_{a=0}^3\left[
- \frac{1}{4} f_{\mu\nu}^a f^{a\mu\nu} + \frac{1}{2} w^a_\mu w^{a\mu}
+ \frac{1}{2}\p_\mu h^a \p^\mu h^a - \frac{\lambda^2}{2}(h^a)^2
\right]
\eeq
with $f_{\mu\nu}^a \equiv \p_\mu w^a_\nu - \p_\nu w^a_\mu$. We also take into
account the external source terms to realize the point vortex
\beq
{\cal L}_{\rm source} = \sum_{a=0}^3 \left[ \rho^a h^a - j_\mu^a w^{a\mu} \right]. \label{sources}
\eeq
The scalar and the vector sources should be determined so that the asymptotic behavior of the fields in
Eq.~(\ref{eq:asymW}) are replicated. The solution of the equation of motion is
\beq
h^0=h^3 = \frac{q}{2\pi} K_0(\lambda r),&&\
\nonumber \rho^0=\rho^3 = q \delta(r),\\
{\bf w}^0={\bf w}^3 = - \frac{m}{2\pi} \hat{\bf k} \times \nabla K_0(r),&&\ {\bf j}^0 = {\bf j}^3= - m \hat{\bf k}
\times\nabla \delta(r)\label{tobedoubled}
\eeq
where $\hat{\bf k}$ is a spatial fictitious unit vector along the vortex world-volume. The vortex configuration with
general orientation is also treated easily, since the origin of the orientation is the Nambu-Goldstone mode associated
with the broken $SU(2)$ color-flavor symmetry
$
H_0 \to H_0(z)^{(1,0)} U_{\rm F},\ \vec\phi_2 = U_{\rm F}^\dagger \vec\phi_1^{(1,0)}$.
The interaction between a vortex at ${\bf x} = {\bf x}_1$ with the orientation $\vec\phi_1$ and another vortex at
${\bf x} = {\bf x}_2$ with the orientation $\vec\phi_2$ is given through the
source term and is summarized as
\beq
V_{\rm int} = - \frac{\left|{\vec \phi}^\dagger_1{\vec \phi}_2\right|^2}{\left|{\vec \phi}_1\right|^2\left|{\vec
\phi}_2\right|^2} \left( \frac{q^2}{2\pi} K_0(\lambda r) -\frac{m^2}{2\pi} K_0(r) \right), \label{eq:eff_e=g}
\eeq
where $r \equiv |{\bf x}_1 - {\bf x}_2| \gg 1$.
When two vortices have parallel orientations, this potential becomes that of two ANO
vortices \cite{Speight:1996px}. On the other hand, the potential vanishes when their orientations are
anti-parallel. This agrees with the numerical result found in the previous sections. In the BPS limit
$\lambda=1$ ($q=m$), the interaction becomes precisely zero.

\subsection{Vortices with general couplings}

It is quite straightforward to generalize the results
of the previous section to the case of generic couplings.
We find
the total potential $V_{\rm int}$
\beq
V_{\rm int} &=& \frac{1}{2}\left(
-\frac{(q^0)^2}{2\pi} K_0(\lambda_e r)
+\frac{(m^0)^2}{2\pi} K_0(r) \right)\nonumber\\
&+& \left(\frac{\left|{\vec \phi}^\dagger_1{\vec \phi}_2\right|^2}{\left|{\vec \phi}_1\right|^2\left|{\vec
\phi}_2\right|^2} - \frac{1}{2}\right) \left( -\frac{(q^3)^2}{2\pi} K_0(\lambda_g \gamma r)
+\frac{(m^3)^2}{2\pi} K_0(\gamma r) \right).
\eeq
At large distance, the interactions between vortices are dominated by the particles with the lowest mass $M_{\rm low}$.
There are four possible regimes $V_{\rm int} =$
\beq
\left\{
\begin{array}{ccll}
-  \frac{(q^0)^2}{4\pi} \sqrt{\frac{\pi}{2 \lambda_e r}}
 e^{- \lambda_e r}  & {\rm for} & M_{\rm low}=M_{\rm s},& {\rm Type} \, {\rm I}\\
 - \left(\frac{\left|{\vec \phi}^\dagger_1{\vec \phi}_2\right|^2}{\left|{\vec \phi}_1\right|^2\left|{\vec
\phi}_2\right|^2} - \frac{1}{2}\right)
   \frac{(q^3)^2}{2\pi} \sqrt{\frac{\pi}{2\lambda_g \gamma r}}
 e^{-  \lambda_g \gamma r}  & {\rm for} & M_{\rm low}=M_{\rm ad},& {\rm Type} \, {\rm I}^* \\
 \frac{(m^0)^2}{4\pi} \sqrt{\frac{\pi}{2 r}}
 e^{- r} & {\rm for} & M_{\rm low}=M_{U(1)},& {\rm Type} \, {\rm II}\\
 \left(\frac{\left|{\vec \phi}^\dagger_1{\vec \phi}_2\right|^2}{\left|{\vec \phi}_1\right|^2\left|{\vec
\phi}_2\right|^2} - \frac{1}{2}\right)
 \frac{(m^3)^2}{2\pi} \sqrt{\frac{\pi}{2 \gamma r}}
  e^{- \gamma r}  & {\rm for} & M_{\rm low}=M_{SU(2)},&{\rm Type} \, {\rm II}^*
\end{array}
\right. ,\label{classif2}
\eeq
because of $K_0(\lambda r)
\sim \sqrt{\pi/2\lambda r} e^{-\lambda r}$.
This generalizes the type I/II
classification of Abelian superconductors. We have found two new categories, called type I$^*$ and type II$^*$, in
which the force can be attractive or repulsive depending on the relative orientation. 
In the type I$^*$ case the forces between parallel vortices are
attractive while anti-parallel vortices repel each other. The type II$^*$ vortices feel opposite forces to the type
I$^*$.  
The result in Eq.~(\ref{classif2}) is easily extended to the general case of $U(1) \times SU(N)$.
This can be done by just thinking of the orientation vectors $\vec \phi$ as taking values
in ${\bf C}P^{N-1}$.

It may be interesting to compare these results with the recently
studied asymptotic interactions between non-BPS non-Abelian global vortices \cite{global}.


\end{document}